\documentclass[letterpaper,twocolumn,10pt]{article}
\usepackage{usenix}

\usepackage{graphicx}
\usepackage{amsmath,amssymb,amsfonts}
\usepackage{algorithm,algpseudocode}
\usepackage{placeins}
\usepackage{enumitem}
\setlist[itemize]{leftmargin=4mm}
\usepackage{xcolor}
\usepackage{xspace}
\usepackage{booktabs}
\usepackage{array}
\newcolumntype{L}[1]{>{\raggedright\arraybackslash}p{#1}}

\DeclareRobustCommand{\sys}{\textsc{PRGuard}\xspace}
\DeclareRobustCommand{\bench}{\textsc{MalPR-Bench}\xspace}

\DeclareRobustCommand{\gap}{VD gap\xspace}

\usepackage{listings}
\definecolor{cCmt}{HTML}{2E7D32}
\definecolor{cKey}{HTML}{0B5394}
\definecolor{cStr}{HTML}{8E24AA}
\definecolor{cHi}{HTML}{FFF2CC}
\lstdefinestyle{prg}{
  basicstyle=\ttfamily\scriptsize,
  keywordstyle=\color{cKey}\bfseries,
  commentstyle=\color{cCmt}\itshape,
  stringstyle=\color{cStr},
  showstringspaces=false,
  breaklines=true,
  breakatwhitespace=true,
  columns=fullflexible,
  keepspaces=true,
  numbers=none,
  xleftmargin=0pt,
  aboveskip=2pt, belowskip=2pt,
  escapeinside={(*}{*)},
}
\lstdefinelanguage{ts}{
  morekeywords={async,await,await,class,const,let,return,if,new,this,export,
                import,from,function,interface,type,public,private,readonly},
  morecomment=[l]{//},
  morecomment=[s]{/*}{*/},
  morestring=[b]',
  morestring=[b]",
  morestring=[b]`,
}
\usepackage{tikz}
\usetikzlibrary{arrows.meta,positioning,fit,backgrounds,calc}
\usepackage{pgfplots}
\pgfplotsset{compat=1.18}
\usepgfplotslibrary{groupplots}
\definecolor{cGPT}{HTML}{0072B2}
\definecolor{cDS}{HTML}{E69F00}
\definecolor{cKB}{HTML}{009E73}
\definecolor{cDet}{HTML}{56B4E9}
\definecolor{cLLM}{HTML}{E69F00}

\begin{document}
\date{}
\title{\Large \bf From Verdict to Diagnosis: Attributable Security Review of Pull Requests}

\author{
{\rm Zhuo Chen}\\
University of Bristol
\and
{\rm Boyang Wang}\\
University of Bristol
\and
{\rm Xiyue Zhang}\\
University of Bristol
\and
{\rm Xiaoyun Xu}\\
Zhongguancun Academy
\and
{\rm Ahmad-Reza Sadeghi}\\
Technical University Darmstadt
\and
{\rm Stjepan Picek}\\
University of Zagreb \& Radboud University
\and
{\rm Lichao Wu}\\
University of Bristol
} %

\maketitle

\begin{abstract}
Automated code reviewers are commonly evaluated by whether they block a malicious pull request (PR). A block, however, may be triggered by an unrelated issue rather than the vulnerability that makes the PR unsafe. Such a review appears successful under verdict-based evaluation, yet can misdirect remediation and leave the actual vulnerability unaddressed. We call this discrepancy between the verdict and the delivered diagnosis the \emph{Verdict--Diagnosis (VD) gap}.

Measuring this gap requires knowing not only whether a PR is vulnerable, but also which vulnerability should be identified and what evidence establishes it. We therefore introduce \bench, a benchmark comprising 89 malicious PRs and 50 benign controls across 44 repositories and eight language families. Its pre-committed, mechanism-level ground truth enables verdict correctness, vulnerability identification, and evidence validation to be evaluated separately.

We further introduce \sys, an attributable PR security reviewer that separates vulnerability identification from evidence validation and retrieves security-relevant repository context to substantiate its findings. Across 31 held-out malicious PRs, \sys and CodeRabbit~\cite{coderabbit}, a widely deployed commercial AI code reviewer, achieve nearly identical blocking performance, yet \sys/DeepSeek identifies 1.38$\times$ as many target vulnerabilities. On absence-type PRs, where a required guard or ownership check is missing, \sys identifies 3$\times$ more vulnerabilities than CodeRabbit. Applied to real-world production repositories, \sys further uncovers twelve previously undisclosed, proof-of-concept-backed vulnerabilities across five widely used projects.
\end{abstract}

\section{Introduction}
\label{sec:intro}

Large language models (LLMs) are increasingly used to generate production
code, yet their output can contain security weaknesses that lead to exploitable
vulnerabilities~\cite{sven}. Autonomous coding agents compound this problem by turning generated code directly into pull requests, scaling both the number of potentially insecure changes and the workload required to assess them~\cite{secdebt}. This automation operates within a rapidly growing review workload: GitHub reports that monthly commit volume more than doubled from 1.4 billion to 2.9 billion between April and August 2026, while merged-PR volume approached 130 million per month~\cite{fedorov2026githuboutage}. Automated reviewers and static-analysis tools thus increasingly mediate which changes receive human attention~\cite{aicodereview2026,cicdauto}. General-purpose code review usually emphasizes whether a change behaves as intended and is maintainable. Security review must additionally assess whether the change creates or leaves open an exploitable path in the surrounding system. Since that path may extend beyond the diff, the decisive evidence can lie in unchanged repository code.

This distinction raises a foundational evaluation question: \emph{what does it mean for an automated PR security review to be correct?} Existing vulnerability-detection evaluations commonly measure whether a system correctly classifies code as vulnerable or benign~\cite{jitvul}. Recent PR-security benchmarks similarly emphasize whether a reviewer approves or rejects a malicious change, sometimes additionally checking whether the rejection raises a security concern~\cite{melo2026sevra}. Still, these metrics do not necessarily establish that the reviewer identified the vulnerability that actually makes the PR unsafe. A correct blocking verdict can therefore conceal an incorrect diagnosis.

We call this discrepancy between a reviewer's verdict and its diagnosis of the actual vulnerability the \emph{Verdict--Diagnosis (VD) gap}. Consider a production PR, detailed in Section~\ref{sec:discovery_vd}
and Table~\ref{tab:discovery}. The PR accepted a client-supplied run identifier, but an unchanged checkpoint lookup did not bind that identifier to the current project, enabling cross-project access. A reviewer blocked the PR, but reported a client-side secret-rendering issue rather than the missing ownership check. Verdict-only scoring would mark this review as successful, even though fixing the reported issue would leave the authorization flaw unaddressed.

This gap motivates our first research question:
\textbf{RQ1: How faithfully does a correct PR-review verdict reflect a correct diagnosis of the underlying vulnerability?} Answering this question requires richer ground truth than a vulnerable/benign label. For each malicious PR, an evaluation must specify not only that the PR is vulnerable but also which vulnerability should be identified and what repository evidence establishes it.
We therefore introduce \bench, a PR-security benchmark to distinguish verdict correctness from both target-vulnerability identification and repository-evidence validation in the delivered review. Each benchmark case has a frozen case-specific rubric specifying the target vulnerability, accepted descriptions of its mechanism, the code evidence required to substantiate it, and off-target security findings that do not receive credit. This allows us to measure whether a reviewer reaches the correct verdict, identifies the target vulnerability, and substantiates that vulnerability with concrete repository evidence.

However, measuring the VD gap alone does not improve the reliability of the security review. We thus ask a follow-up question: 
\textbf{RQ2: Can separating vulnerability identification from evidence validation help PR reviewers narrow the VD gap?}
To answer this question, we propose \sys, which first formulates a concrete
candidate vulnerability and then tests its critical premises against repository
evidence. Since evidence may lie beyond the changed lines, \sys examines
the repository context surrounding the change and retrieves additional evidence
when needed to resolve a concrete security question. We evaluate how target-vulnerability identification and evidence validation vary across configurations that provide different forms and amounts of repository evidence. The goal of \sys is therefore not merely to block more malicious PRs, but to 
make each blocking verdict attributable to the actual vulnerability introduced by the pull request and grounded in repository evidence that substantiates the diagnosis.

Controlled benchmarks measure the VD gap against vulnerabilities known to the
evaluators, but real-world review begins without a predefined target or rubric.
We therefore ask:
\textbf{RQ3: Can \sys uncover real-world vulnerabilities?}
We apply \sys to historical PR candidates from production repositories and
confirm twelve previously undisclosed, PoC-backed vulnerabilities across five
repositories. On these cases, independent runs of \sys and CodeRabbit both
block 10/12 PRs, but produce 10/12 and 4/12 attributable blocks, respectively.
Thus, identical verdict totals mask a 2.5$\times$ difference in attributable
blocks.
Our contributions are as follows:
\begin{itemize}
\itemsep0em 
\item We formulate and characterize the \emph{Verdict--Diagnosis gap} in automated PR security review and introduce a systematic evaluation framework that individually measures verdict correctness, target-vulnerability identification, and evidence validation.

\item We introduce \bench, a mechanism-grounded benchmark comprising 89 malicious PRs and 50 benign controls across 44 repositories and eight languages. Its pre-committed rubrics specify both the target vulnerability and the repository evidence to support it.

\item We design \sys, an attributable PR security reviewer that separates vulnerability identification from evidence validation to bridge verdicts with evidence.

\item On the 31-case common-coverage challenge set, \sys/DeepSeek identifies 1.38$\times$ as many target vulnerabilities as CodeRabbit despite similar blocking totals. On absence-type PRs, both systems block 9/14 cases, but \sys/DeepSeek identifies 3$\times$ more vulnerabilities. We further establish the practical significance of the VD gap by uncovering twelve previously undisclosed and PoC-backed vulnerabilities across five production repositories.
\end{itemize}

The paper is organized as follows: Section~\ref{sec:prelim} defines the threat model and attributable security review problem. Section~\ref{sec:bench} presents \bench and its mechanism-level evaluation methodology. Section~\ref{sec:design} introduces the design and implementation of \sys. Section~\ref{sec:results} evaluates the Verdict--Diagnosis gap and \sys on held-out PRs from \bench. Section~\ref{sec:discovery} studies previously undisclosed production vulnerabilities. Section~\ref{sec:discussion} discusses the implications and scope of the findings, and Section~\ref{sec:related} reviews related work. Finally, Section~\ref{sec:conclusion} concludes this work. The appendices provide construction and reproducibility records, case-level
evidence, implementation and grading details, ethical considerations, and
open-science documentation.

\section{Problem Definition}
\label{sec:prelim}

In the client-supplied identifier case introduced in Section~\ref{sec:intro}, the reviewer reached the correct blocking verdict but failed to diagnose the resulting cross-project authorization vulnerability. To make this distinction precise, in this section, we specify what a PR reviewer observes, what the PR author controls, and what a successful review must deliver.

\subsection{PR Security Review and Threat Model}
\label{sec:threat}

A pull request (PR) proposes a set of code changes against a fixed repository revision, together with metadata such as its title and description. A PR security reviewer examines the proposed change and relevant repository context and returns a review containing a \emph{verdict} and, when applicable, one or more security findings. We focus on PR-level security review: deciding whether the proposed change is safe to merge and explaining any vulnerability newly introduced by the PR or left exploitable by an incomplete fix. Full-repository vulnerability auditing and exploit generation are out of scope.

For each malicious PR, we designate the security defect that the reviewer is expected to identify as the \emph{target vulnerability}. A benign PR neither introduces a target vulnerability nor leaves one exploitable. When a paired benign control is available, it contains the complete fix for the corresponding malicious case, while unrelated security issues may still exist elsewhere in the repository. In rare cases, a PR may contain multiple security issues. Each benchmark case designates one as the target vulnerability; alternative valid descriptions and evidence chains are handled by the case-specific rubric in Section~\ref{sec:bench_design}.
The target vulnerability need not be visible in the diff; locating it may require inspecting unchanged security-relevant code elsewhere in the repository.

We use a conservative adversarial model in which the PR author controls the proposed diff, PR title and description, and all repository content added or modified by the PR, including comments, documentation, and tests. This model covers untrusted external contributions, compromised developer accounts, and agent-generated PRs; it does not assume that ordinary contributors are malicious. The pre-PR repository revision and the review infrastructure remain within the defender's trust boundary. As the reviewer cannot infer the author's intent, it treats all PR-controlled material as untrusted: comments and documentation may assert that a security condition holds, and tests may encode the same assumption, but such claims must be verified against the implementation and surrounding repository context.
The attacker's objective is to introduce a vulnerability or leave one unresolved without having the review clearly identify it. The defender must therefore answer two questions: Is the PR safe to merge? If not, what vulnerability must be fixed? These questions correspond to the verdict and diagnosis defined next.

\subsection{Verdict--Diagnosis Gap}
\label{sec:vdgap}

A PR security review produces two observable outputs. The \emph{verdict} states whether the reviewer considers the PR safe to merge. The \emph{diagnosis} states what security problem, if any, justifies that verdict. These outputs answer different questions and must therefore be evaluated separately.

For a malicious PR, a correct diagnosis must first answer: \emph{what is the vulnerability?} We call this requirement \emph{vulnerability identification}. The review must characterize the target concretely by stating the affected behavior, the missing or violated security condition, and the resulting consequence. A broad category such as ``authorization issue'' is insufficient because it neither distinguishes the target from neighboring security concerns nor tells the maintainer which behavior must change. Acting on such a diagnosis may therefore leave the target vulnerability exploitable.

On the other hand, a correct diagnosis must also answer: \emph{what repository evidence establishes this vulnerability?} We call this requirement \emph{evidence validation}. A reviewer may name a plausible vulnerability mechanism without establishing that its required premises hold in the current repository. Evidence validation distinguishes such speculation from an established finding by grounding the security-critical premises in concrete, auditable repository facts and code locations. The required premises are case-specific: they may concern attacker control, the need for a missing guard, a relevant state transition, or a security-sensitive operation.

Returning to the production case introduced in Section~\ref{sec:intro}, identification requires stating that the client-supplied run identifier reaches a checkpoint lookup that is not scoped to the current project, enabling cross-project access. Evidence validation additionally requires grounding the client-controlled identifier, the unscoped lookup, and repository evidence showing that the lookup should be restricted to the current project. The reported client-side secret-rendering issue does not identify or validate the target vulnerability.

A PR block is \emph{attributable} when its delivered diagnosis both identifies and substantiates the target vulnerability. The \emph{Verdict--Diagnosis (VD) gap} occurs when verdict correctness and diagnosis correctness diverge. Its central failure mode is a correct blocking verdict accompanied by an off-target or unsubstantiated diagnosis. Conversely, a reviewer may correctly identify the target vulnerability without assigning it a blocking verdict. Section~\ref{sec:bench_scoring} operationalizes these distinctions as \(V\), \(I\), \(E\), and \(A\).

\section{\bench}
\label{sec:bench}

Section~\ref{sec:vdgap} defines an attributable review, but a definition alone
does not make the VD gap measurable. The remaining challenge is to turn its
verdict and diagnosis requirements into case-level ground truth that can be
applied consistently across reviewers. \bench provides this measurement layer.
We first specify the ground truth and scoring procedure, then explain how cases
are constructed and annotated to reveal where diagnosis fails.

\subsection{Measurement Requirements}
\label{sec:bench_design}

For each malicious PR, \bench fixes three case-specific judgments: \emph{whether the PR should be blocked}, \emph{which vulnerability justifies that disposition}, and \emph{which security-critical premises a correct diagnosis must establish}. Each case pins the base repository revision, review-visible PR metadata, and proposed diff, and the delivered review is scored against this fixed contract.

The target-vulnerability field defines what the review must identify; the
validation requirements define what it must establish. The grading boundary admits semantically equivalent descriptions while excluding off-target
diagnoses. Before grading begins, the case rubric is frozen and subsequently applied unchanged to every reviewer. If a rubric is subsequently revised, all dependent grades are invalidated and recomputed. A PR may contain multiple security issues or admit more than one valid evidence chain. In such cases, the frozen rubric records the target mechanism together with alternative descriptions or evidence chains supported by the repository. A delivered review receives credit if it satisfies an accepted path; unrelated findings do not substitute for the target vulnerability.

When a corresponding secure state is available, we pair the malicious PR with a benign control that contains the complete fix while preserving the surrounding repository context. The control tests whether a reviewer distinguishes the vulnerable state from its corrected counterpart rather than blocking code merely because it is security-sensitive. Each control is reviewed and scored independently.
This measurement contract specifies the reference against which a delivered
review is judged. We next describe how the contract is converted into
observable review scores.
\begin{table}[t]
    \centering
    \small
    \setlength{\tabcolsep}{4pt}
    \caption{Case-level ground truth for scoring a malicious PR.}
    \label{tab:case-contract}
    \begin{tabular}{@{}lL{0.62\columnwidth}@{}}
        \toprule
        Field & Purpose \\
        \midrule
        Expected verdict
        & Whether the PR should be blocked (\(V\)). \\

        Target vulnerability
        & Affected behavior, missing or violated security condition, and consequence (\(I\)). \\

        Validation requirements
        & Security-critical premises that the diagnosis must ground in repository evidence (\(E\)). \\

        Grading boundary
        & Accepted equivalent descriptions and repository-supported evidence chains; excluded off-target findings. \\
        \bottomrule
    \end{tabular}
\end{table}

\subsection{Operational Scoring and Grading}
\label{sec:bench_scoring}

Given a frozen case rubric and a delivered review, \bench assigns three binary component scores. For a malicious PR, \(V=1\) when the reviewer returns a blocking verdict. We assign \(I=1\) when the delivered diagnosis correctly characterizes the target vulnerability, including the affected behavior, the missing or violated security condition, and the resulting consequence. We assign \(E=1\) when \(I=1\) and the delivered review grounds the rubric-specified security-critical premises in correct repository facts and code locations.
A review constitutes an \emph{attributable block} when the PR is blocked; the diagnosis identifies and substantiates the target
vulnerability:
\[A = V \land I \land E.\]
The \(V\), \(I\), and \(E\) scores expose where verdict and diagnosis
diverge, while \(A\) records whether the blocking verdict is attributable to
the vulnerability that justifies it.
Both \(I\) and \(E\) are properties of the delivered review. Scoring them does
not require access to a reviewer's internal candidate-generation or validation
process.
A benign control has no target vulnerability to identify. We thus score
whether it is incorrectly blocked, recording a blocking verdict as a false
positive.

Verdict scoring is deterministic from the delivered disposition. Two LLM graders independently proposed \(I\) and \(E\) labels using the frozen rubric and delivered review, blinded to system identity, backend, pipeline configuration, and aggregate results. Two human graders jointly adjudicated all disagreements between the LLM graders and jointly spot-checked a subset of their agreements; the resulting joint decision determined the final label. Because the human graders adjudicated jointly rather than producing independent label sets, a human inter-rater statistic is not defined for this workflow.

\subsection{Benchmark Construction}
\label{sec:bench_composition}

The scoring scheme above defines how a fixed PR is evaluated. We instantiate it using cases from three complementary sources: mined project histories, public security advisories, and validated production discoveries. These sources differ in how cases are selected and which experimental claims they support. We therefore retain separate tiers rather than treating all cases as a homogeneous test set.

\noindent\textbf{Mined history.}
We recover \emph{incomplete-fix residuals} from project histories by identifying later security patches that complete an earlier fix and reconstructing the post-fix, pre-correction repository state, in which a related path remained vulnerable~\cite{szz}.
Fourteen malicious cases from five repositories form the development tier and are used only to construct the mechanism knowledge consumed by \sys. They are not included in the held-out performance results. A separate mined-history tier, which we call \emph{self-generalization}, contains 19 malicious PRs and six benign controls from five repositories. The two tiers use the same history-recovery procedure, but no repository in the
self-generalization tier contributes knowledge to \sys.

\noindent\textbf{Advisory-derived external pools.}
To reduce dependence on our history-mining procedure, we derive two additional pools from GitHub-reviewed advisories published in 2025~\cite{githubAdvisoryDatabase}. Fix commits and their parent revisions are resolved following the CVEfixes methodology~\cite{cvefixes}, and cases are admitted under fixed eligibility and quality criteria before any evaluated reviewer is run.
Pool~A models an incomplete fix by retaining the production changes of a real security patch while withholding one security-enforcement change. Pool~B reverses a localized enforcement change, placing the removed protection directly in the reviewed diff. In both pools, the complete real fix forms the paired benign control. These cases are controlled review inputs derived from disclosed fixes; we do not claim that the resulting malicious states were deployed by the original projects. The construction yields seven Pool~A pairs and 37 Pool~B pairs.
Pool~A and Pool~B expose complementary review conditions. Pool~B makes the
relevant enforcement removal visible in the diff, whereas Pool~A requires the reviewer to recognize that required enforcement is absent. The complete eligibility criteria and selection accounting appear in Appendix~\ref{app:external-construction} and the shared artifact.

\noindent\textbf{Real-world discovery challenge set.}
To make the \bench reflect real-world scenarios, \bench additionally includes a twelve-case production-discovery challenge set comprising previously undisclosed, PoC-backed vulnerabilities discovered by the \sys across five production repositories (detailed in Section~\ref{sec:discovery}). None of these repositories contributes knowledge to \sys. After independent validation, we froze each repository revision, target vulnerability, and case rubric before running \sys and CodeRabbit. Since the originating \sys configuration selected these cases, its performance on this tier is selection-biased and is reported only for context. The independently run reviews test whether the VD gap also appears on validated production vulnerabilities; this tier does not provide an unbiased estimate of the originating configuration's performance.

Table~\ref{tab:bench} summarizes the resulting benchmark. Out of its 89 malicious
cases, 14 are reserved exclusively for knowledge-base development. The
remaining 75 come from repositories that contribute no knowledge to \sys.
The benchmark additionally contains 50 paired benign controls. Results are
reported by tier because the controlled, mined, and discovery cases have
different selection effects and experimental roles.

\begin{table}[t]
    \centering
    \caption{\bench composition. The 14 development cases are used only for knowledge-base construction; the remaining 75 are held out from knowledge-base construction, although their selection effects and experimental roles differ by tier.
    Repositories in held-out tiers contribute no knowledge-base entry.}
    \label{tab:bench}
    \begin{tabular}{l l r r r}
        \toprule
        Tier & Origin & Repos & Malicious & Benign \\
        \midrule
        Develop. & Mined hist. & 5 & 14 & -- \\
        Self-gen. & Mined hist. & 5 & 19 & 6 \\
        Pool A & Advisory & 7 & 7 & 7 \\
        Pool B & Advisory & 22 & 37 & 37 \\
        Discovery & PRGuard & 5 & 12 & -- \\
        \midrule
        Total & -- & 44 & 89 & 50 \\
        \bottomrule
    \end{tabular}
\end{table}

\subsection{Diagnostic Case Properties}
\label{sec:bench_properties}

To understand why a PR diagnosis fails, \bench annotates each malicious case along
two properties. \emph{Defect class} describes what a correct validation must
establish, while \emph{evidence location} describes how far beyond the diff a
reviewer must look to obtain the required repository facts. Both annotations
are fixed from the case rubric and repository revision before any reviewer is
evaluated.

\noindent\textbf{Defect class.}
We distinguish \emph{present-type} and \emph{absence-type} defects. A
present-type defect can be validated from security-relevant behavior that is
present in the reviewed evidence, such as an unsafe operation, an insecure
state transition, or an authorization check explicitly removed by the PR. An
absence-type defect instead requires establishing that required security enforcement is missing, such as an ownership check, validation guard, or
required state update. Since the missing operation has no line of its own to cite, validation must establish both the unguarded path and why the omitted enforcement is required, for example, a guarded peer, a sibling path performing the same sensitive operation that does apply the check, or a repository contract, an interface, invariant, or documented convention that mandates it. The classification follows this
validation requirement: deleting an existing guard
is present-type, whereas failing to add a required guard is absence-type.

\noindent\textbf{Evidence location.}
Let \(C\) denote the changed lines in the PR diff and \(T\) the files touched by the PR; we consider four evidence locations:
\begin{itemize}
    \itemsep0em 
    \item \textbf{L0} indicates that the validation requirements can be satisfied from \(C\) alone.
    \item \textbf{L1} additionally requires unchanged code elsewhere in \(T\).
    \item \textbf{L2a} requires code outside \(T\) connected to the change through a structural relation, such as a call or an import.
    \item \textbf{L2b} requires code outside \(T\) connected to the change by semantic correspondence, such as a sibling endpoint or equivalent handler.
\end{itemize}
When validation requires facts from multiple locations, the case receives the most non-local applicable label. Appendix~\ref{app:evidence-loc} presents case-level annotations for the self-generalization tier.

Overall, defect class and evidence location capture distinct case properties: the former describes the validation obligation, whereas the latter describes the required repository scope. We therefore analyze them separately rather than combining them into a single difficulty label.

\section{\sys}
\label{sec:design}
Having established how the VD gap can be measured, we now turn from measurement to mitigation. To narrow the VD gap, \sys decomposes PR security review into a sequence of
stages that separately characterize the change, acquire repository evidence,
construct and validate candidate vulnerabilities, and produce an attributable
final review.
\begin{figure*}[t]
    \centering
    \includegraphics[width=\textwidth]{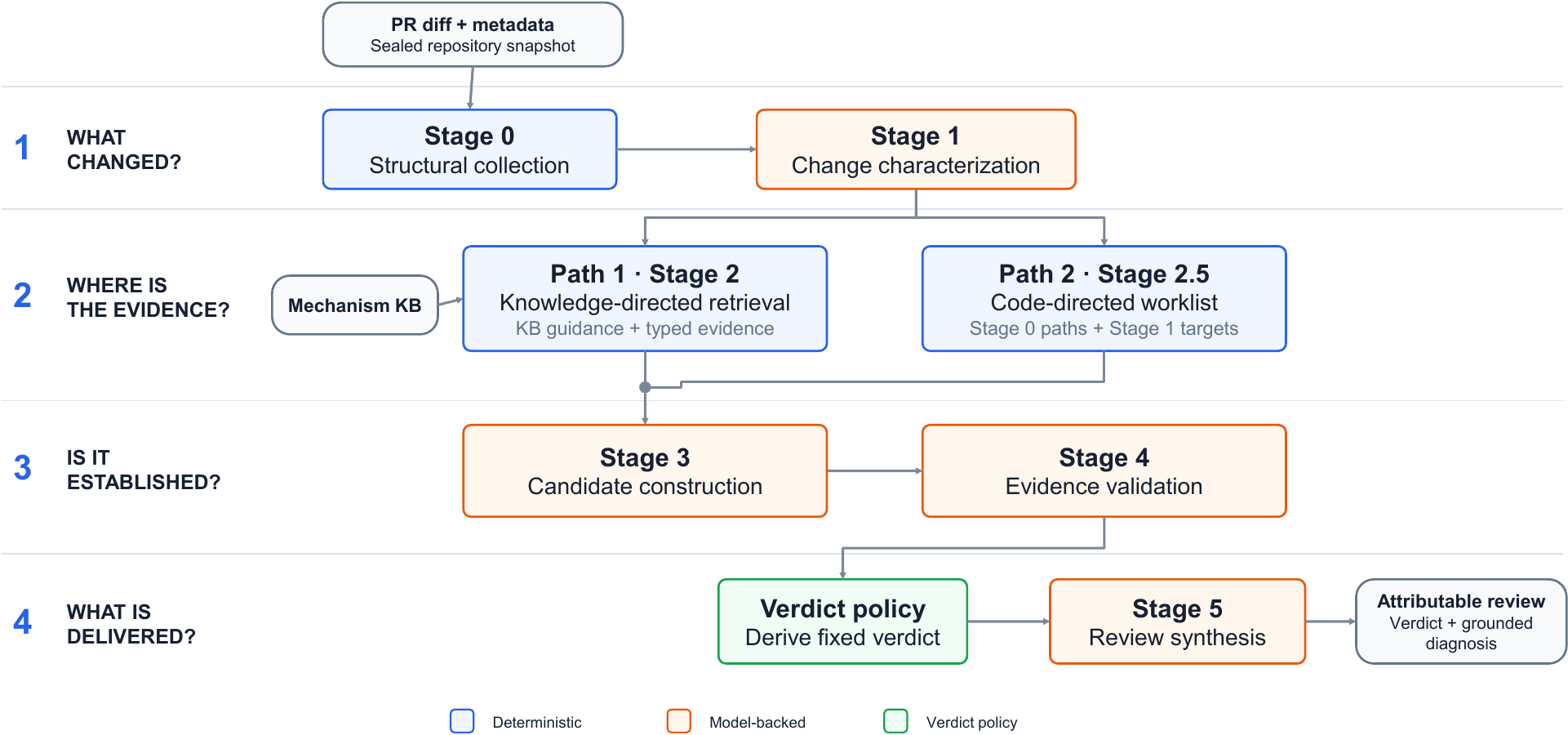}
    \caption{\sys decomposes PR security review around four questions.
    Structural collection and knowledge-directed retrieval acquire
    current-repository evidence; separate identification and validation stages
    distinguish a plausible candidate from an established vulnerability. A
    deterministic policy derives the verdict, after which review synthesis
    explains the validated result.}
    \label{fig:pipeline}
\end{figure*}

\subsection{Design Overview}
\label{sec:design_overview}
A PR diff shows where code changed, but it may not reveal which security
condition is missing, or where the evidence needed to establish a vulnerability
resides. Jumping directly from the diff to a verdict therefore risks allowing
an early hypothesis to determine both what evidence is retrieved and how that
evidence is interpreted. \sys instead turns the review into four successive
questions. As shown in Figure~\ref{fig:pipeline}, \sys first asks \emph{what changed}, without yet proposing a vulnerability:
Stage~0 deterministically collects structural context around the change, and
Stage~1 characterizes its security-relevant behavior and unresolved questions.
Once the change is characterized, evidence acquisition splits into two complementary paths. Path 1 (knowledge-directed): Stage 2 uses mechanism knowledge to retrieve repository evidence through a typed relation. Path 2 (code-directed): Stage 2.5 builds a KB-independent worklist from the changed code and its structural context.
This separation lets
the KB guide where to look without treating stored knowledge as proof or
allowing an irrelevant match to suppress code-directed analysis. The two paths
converge only after evidence has been collected: Stage~3 constructs explicit
candidate vulnerabilities, and Stage~4 tests the same candidates against
attributable repository evidence. Finally, a deterministic policy maps the
validation outcomes to a verdict, and Stage~5 renders that fixed result as the
review delivered to the maintainer.

\subsection{What Did the PR Change?}
\label{sec:structural_context}

\noindent\textbf{Stage~0: Evidence-first structural collection.}
Stage~0 runs before any model reasoning, so that the model's first hypothesis
does not determine which repository code later stages can see. Starting from
the changed functions, fixed static collectors retrieve three kinds of context:
direct structural connections such as callers, callees, definitions, imports,
and contracts; cross-location data uses, such as code that produces or consumes
the same value or context key; and security-coverage comparisons, such as other
call sites reaching the same sensitive operation or peer handlers expected to
apply the same guard. The retrieved locations and code excerpts form a bounded
structural context pack. The collectors are fixed across PRs and do not use the
KB.

\noindent\textbf{Stage~1: Change characterization.}
Stage~0 organizes context around code relationships, but it does not explain
which of those relationships matter to the behavior changed by the PR.
Stage~1 uses the review model to convert the diff and structural context pack
into a security-oriented description of the change. It identifies the affected
components, describes their behavior before and after the change, and records
the relevant source, changed behavior, guard or state, sensitive sink, and
security effect. When the supplied code does not establish one of these
elements, Stage~1 records it as unknown rather than completing the chain by
assumption.

The resulting change characterization serves two purposes. Its structured
mechanism summary becomes the retrieval query used by Stage~2, while its
touched components and unresolved review questions become the code-directed
targets used by Stage~2.5. Stage~1 receives no knowledge-base entry and does not propose a candidate vulnerability or assign a verdict.

\subsection{Where Is the Required Evidence?}
\label{sec:evidence_acquisition}

\noindent\textbf{Mechanism KB.}
We manually derive the KB using only the fourteen development cases. With
Stage~2 disabled, we classify each rubric failure as missing evidence or
missing reasoning. Evidence failures yield either a reusable Stage~0 collection
rule or a KB entry specifying a retrieval anchor and typed relation; reasoning
failures yield guidance for Stage~3 or Stage~4. We retain a change only if it
fixes the originating case without regressing previously successful development
cases, tested by forcing the proposed entry into the other cases. Entries
encoding the same security condition and retrieval relation are merged.
Retrieval parameters follow the same development-only regression process, and
the KB and configuration are frozen before held-out evaluation and released
with their derivation records. Because the KB is derived from these cases, we treat development-set improvement as motivation and rely on the held-out tiers for effectiveness.
The frozen KB contains 21 mechanisms represented by 25 entries. Appendix~\ref{app:implementation} presents its
stage-target and retrieval-mode.

\noindent\textbf{Stage~2: mechanism matching and conditional typed retrieval.}
Stage~2 forms a query from the non-unknown fields of the Stage~1 mechanism
summary, i.e., source, changed behavior, guard or state, sink, and security
effect, together with its change summary and touched components. It compares
this query with each KB trigger using semantic similarity and retains at most
five mechanisms above the coarse threshold.

The retained mechanisms follow two retrieval modes. A \texttt{single}
entry requires no further repository fetch and may pass its guidance
forward based on the Stage~1--trigger match. For a \texttt{two\_stage}
entry, that match only nominates the mechanism: its retrieval contract
must also be resolved against the current repository. The contract pairs
an anchor with one of four fixed relations, i.e., \textsc{Caller-Source},
\textsc{Consumer-Sink}, \textsc{Guard-Def}, or
\textsc{Sibling-Endpoint}. The anchor, either named by the entry or
derived from the collected structural context, determines where the
retrieval begins; the relation determines which connected code to
retrieve. Stage~2 returns the resulting locations and excerpts together
with their structural connection to the anchor. A \texttt{two\_stage}
mechanism is ineligible if any requested relation cannot be grounded.

Stage~2 assigns each coarse candidate
\(s_{\mathrm{fine}} =
0.8s_{\mathrm{coarse}} + 0.2s_{\mathrm{structural}}\),
where the structural term is the fraction of requested relations resolved
and is zero for \texttt{single} entries. To bound downstream reasoning,
Stage~2 retains at most three candidates above the fine threshold; both the
threshold and cap are fixed through the development-only regression process
described above. It keeps their guidance separate from any current-repository
code returned by typed retrieval.

Authentik PR \#9076\footnote{\url{https://github.com/goauthentik/authentik/pull/9076}}
illustrates \textsc{Sibling-Endpoint}. The changed OAuth2
device-code flow is not structurally connected to the unchanged authorization
path that invokes \texttt{PolicyAccessView}; the typed sibling relation
retrieves that role-equivalent path as a security comparator.
Appendix~\ref{app:sibling} records the case-level non-isolation evidence.

\noindent\textbf{Stage~2.5: structural review worklist.}
Independently of the knowledge-base match, Stage~2.5 maps the components
identified by Stage~1 to the reference paths collected by Stage~0,
forming a KB-independent worklist of repository paths for subsequent examination. It contains
neither an external vulnerability mechanism nor a verdict, ensuring that an empty or irrelevant Stage~2 match cannot suppress targets derived from the changed code.

\subsection{Is the Vulnerability Established?}
\label{sec:identify_validate}

\noindent\textbf{Stage~3: candidate-vulnerability construction.}
Stage~3 receives the PR and Stage~1 characterization together with the collected structural context, the Stage~2.5 worklist, and any Stage~2 guidance and repository evidence. It formulates each candidate as a concrete explanation of how the PR creates or preserves an exploitable behavior and what security consequence follows.
Each candidate records the mechanism, attacker capability, exploit
sequence, reachability conditions, impact, confidence, supporting
locations, and any missing evidence. An incomplete but plausible chain may be retained at low confidence with its
proof gap stated explicitly, whereas a chain contradicted by the repository is
omitted. If no candidate remains, Stage~4 is skipped. Stage~3 thus proposes
what the vulnerability might be; Stage~4 separately determines whether the
repository evidence establishes it.

\noindent\textbf{Stage~4: evidence validation.}
Stage~4 is a separate model invocation that tests the same candidates against current-repository evidence rather than searching for a different vulnerability. For each candidate, it checks the security-critical premises concerning attacker control, reachability, missing or violated security conditions, sensitive operations, and consequences against attributable repository locations.
A candidate is \textsc{Validated} when the complete chain is supported;
\textsc{Downgraded} when a required link remains incomplete or the
behavior is better explained as a non-security correctness concern; and
\textsc{Rejected} when the chain is contradicted or unsupported. The
structured result records the supporting evidence and any benign
alternative considered. Only \textsc{Validated} candidates are treated
as established vulnerabilities, implementing the evidence-validation
criterion of Section~\ref{sec:bench_scoring}.

\emph{Bounded evidence closure.}
When validation is blocked by a missing definition on the candidate's
existing evidence path, Stage~4 may request that symbol by name. A fixed
resolver retrieves a bounded method or class body from the sealed
repository, reusing an untruncated Stage~0 excerpt when available, and
then reruns validation of the same candidate. Requests, definitions, and
rounds are deduplicated and capped as detailed in
Appendix~\ref{app:implementation}; the procedure cannot introduce a new
vulnerability or initiate free-form repository search.

\subsection{What Should be Delivered?}
\label{sec:verdict_synthesis}

Stage~4 outcomes are mapped to a verdict by a deterministic policy. Any
\textsc{Validated} vulnerability produces \textsc{BLOCK}; if none is
validated but at least one is \textsc{Downgraded}, the result is
\textsc{COMMENT}; otherwise the result is \textsc{APPROVE}.

\noindent\textbf{Stage~5: review synthesis.}
Stage~5 receives the structured validation outcomes and predetermined
verdict and converts them into the review delivered to the maintainer.
It is constrained to explain the supported findings with attributable
repository locations rather than propose a new vulnerability or revise the
verdict.

\subsection{Security Boundary and Implementation}
\label{sec:security_boundary}

As defined in Section~\ref{sec:threat}, PR-controlled repository content
is untrusted. \sys operates on a sealed repository snapshot and a frozen,
read-only knowledge store. It neither executes repository code nor permits
PR-directed network retrieval, and reviewed content cannot modify the
pipeline control graph. Model-directed actions are limited to the four
Stage~2 retrieval relations and the bounded named-definition requests of
Stage~4.
These restrictions constrain how repository content can affect evidence
acquisition, but not how it influences model judgment. Prompt injection
may still bias the Stage~1 characterization, steer later reasoning toward
an irrelevant hypothesis, or consume the available evidence budget. We
therefore treat prompt injection, judgment errors, and budget exhaustion
as residual attack surfaces.

Each run records its repository revision and configuration, structured
stage outputs, selected KB identifiers, retrieved repository evidence,
recovery telemetry, and final verdict, allowing the deterministic
evidence pipeline to be replayed. Stage~0 is deterministic at fixed
inputs, while Stage~2 is deterministic given its saved Stage~0 and
Stage~1 inputs and retrieval configuration. Model-backed stages remain
provider-nondeterministic.
Note that \sys is not hard-coded to a particular backend: another
instruction-following model can adapt to \sys when it satisfies the structured-output format. Detailed prompts, limits, retrieval parameters, and implementation
inventory appear in Appendix~\ref{app:implementation}.

\section{Evaluation}
\label{sec:results}

This section addresses the first two research questions. First, we examine whether verdict-level scoring faithfully reflects the security diagnosis delivered by a reviewer (RQ1). Second, we evaluate \sys at the diagnosis level and use an evidence-configuration ladder to examine how identification and validation differ across configurations (RQ2). Section~\ref{sec:discovery} separately addresses RQ3 by studying previously undisclosed production vulnerabilities.

\subsection{Experimental Setup}
\label{sec:eval_setup}

We evaluate the same \sys pipeline with GPT-5.5 (proprietary) and DeepSeek v4-pro (open-weight) as review backends. Both use the same prompts, retrieval index, stage limits, and deterministic verdict policy described in Section~\ref{sec:design}. Our primary baseline is CodeRabbit~\cite{coderabbit}, a commercial AI code reviewer distributed through the GitHub Marketplace and integrated directly into pull-request workflows. At the time of writing, GitHub Marketplace reports more than 300,000 CodeRabbit installations~\cite{coderabbitMarketplace}, while CodeRabbit reports serving more than 17,000 customers~\cite{coderabbitCustomers}. These adoption figures motivate its selection as a widely deployed product baseline. 

As a black-box product, CodeRabbit does not disclose its model, inference configuration, or intermediate reasoning. We therefore compare only delivered outputs: the results reflect end-to-end product performance, not the effects of particular internal design choices. For CodeRabbit, an actionable Critical or Major security finding maps to \textsc{BLOCK}, and its delivered comments are graded for \(I\) and \(E\) against the same frozen case-specific \bench rubric as \sys. We additionally run CodeQL's stock security-extended suite~\cite{codeql} as a static-analysis reference, on the cases its queries apply to.

\noindent\textbf{Single-run scoring.}
All reported performance scores use one predesignated review per case (Run~1). Verdict
\(V\), vulnerability identification \(I\), and evidence validation \(E\) are
scored from the delivered review under Section~\ref{sec:bench_scoring}; a
benign control is a false positive when that review blocks. A second independent PRGUARD run is used only as a rerun diagnostic over the external pools and does not replace the predesignated Run~1 in any headline or comparative performance claim. CodeQL is
deterministic and is run once.

\noindent\textbf{Evaluation sets.}
Both complete \sys configurations were evaluated on all 75 malicious cases held out from knowledge-base construction and all 50 benign controls. Because of CodeRabbit's rate limits and evaluation cost, we ran it on two complete tiers: all 19 self-generalization cases and all 12 discovery cases. We did not run CodeRabbit on Pools~A or~B, and no case within either reviewed tier was selected based on CodeRabbit's output. Cross-system comparisons therefore use these 31 common-coverage cases.
In these 31 cases, CodeRabbit's \(I\) and \(E\) coincide: every comment identifying the target also supplies the required evidence. This is an observed property of its outputs, not a scoring rule.
The twelve discovery cases were originally surfaced by the \sys. Its score on that tier is therefore selection-biased and is reported only for context. DeepSeek and CodeRabbit were run independently after the cases and rubrics were frozen.

\subsection{Overall Performance}
\label{sec:vd_results}

We first examine the central question of this paper: whether a correct blocking
verdict implies that the reviewer identified the vulnerability that makes the
PR unsafe.

\noindent\textbf{Verdict--diagnosis gap.}
Table~\ref{tab:defect} shows that the verdict--diagnosis gap is concentrated in
absence-type defects, where the diagnosis must recognize missing enforcement.
Across the 31 common-coverage challenge cases, CodeRabbit blocks 24 PRs and
\sys/DeepSeek blocks 22. Their delivered diagnoses reverse this ordering:
\sys/DeepSeek identifies 22 target vulnerabilities, compared with 16 for
CodeRabbit, or 1.38$\times$ as many. On the 14 absence-type cases, both systems block
9 PRs, while \sys/DeepSeek identifies 9 target vulnerabilities and CodeRabbit
identifies 3, an exact threefold difference. Thus similar verdict totals can
conceal a substantial difference in the security problem communicated to the
maintainer.

\begin{table}[t]
    \centering
    \small
    \caption{Performance on the 31 common-coverage challenge cases stratified by defect class.
    \(V\) denotes a blocking verdict and \(I\) target-vulnerability
    identification.}
    \label{tab:defect}
    \begin{tabular}{lrrrr}
        \toprule
        & \multicolumn{2}{c}{Present-type (17)}
        & \multicolumn{2}{c}{Absence-type (14)} \\
        \cmidrule(lr){2-3}\cmidrule(lr){4-5}
        System & \(V\) & \(I\) & \(V\) & \(I\) \\
        \midrule
        \sys/DeepSeek
            & 13/17 & 13/17 & 9/14 & 9/14 \\
        CodeRabbit
            & 15/17 & 13/17 & 9/14 & 3/14 \\
        \bottomrule
    \end{tabular}
\end{table}

Table~\ref{tab:common-attribution} applies the complete output-level scoring
contract to the same two independently run reviewers. Although CodeRabbit's
\(I\) and \(E\) marginals coincide on these cases, attribution is still computed
case by case as \(A=V\land I\land E\), rather than inferred from marginal totals.
For context, the originating \sys/GPT-5.5 configuration blocks 25/31 cases and
identifies 26/31 targets, including all twelve cases in the discovery tier. We
report those values descriptively, not as an unbiased estimate of that
configuration's performance.
\begin{table}[t]
    \centering
    \small
    \caption{Output-level scores for the independently run reviewers on the
    31 common-coverage challenge cases. \(A\) denotes an attributable block.}
    \label{tab:common-attribution}
    \begin{tabular}{lrrrr}
        \toprule
        System & \(V\) & \(I\) & \(E\) & \(A\) \\
        \midrule
        \sys/DeepSeek & 22/31 & 22/31 & 20/31 & 19/31 \\
        CodeRabbit & 24/31 & 16/31 & 16/31 & 16/31 \\
        \bottomrule
    \end{tabular}
\end{table}

Evidence location provides an additional view. CodeRabbit identifies the
target vulnerability in 16/24 L0/L1 cases, but in 0/7 L2a/L2b cases, where
validation requires evidence outside the touched files (one-sided Fisher exact
\(p=0.002\)). These seven cases span different security mechanisms instead of
one recurring vulnerability pattern. We therefore observe a strong association
between outside-file evidence requirements and diagnosis failure, without
claiming that evidence location alone determines reviewer performance.

Stock CodeQL's security-extended queries apply to 21 of the 31 cases. It produces one off-target blocking finding and identifies none of the target vulnerabilities.

\subsection{\sys Assessment}
\label{sec:understanding}

We next evaluate \sys on all 63 held-out malicious cases (19 self-generalization + 7 Pool A + 37 Pool B), excluding the discovery tier (Section 6) because those cases were surfaced by PRGuard and are therefore selection-biased.

\noindent\textbf{Tier-wise complete-pipeline performance.}
Table~\ref{tab:complete} reports \(V\), \(I\), \(E\), and \(A\) for the complete pipeline on 63 held-out cases: 19 self-generalization cases and 44 cases from Pools~A and~B. Unlike the 31-case comparison, this evaluation excludes the discovery tier, and no evaluated repository contributed to the knowledge base. The metrics are marginal rather than sequential: \(V\) scores the final disposition, whereas \(I\) and \(E\) score the delivered review; thus \(I\) may exceed \(V\), while \(E=1\) requires \(I=1\).

Since Pool~B contributes 37 of the 63 cases, the pooled row is descriptive rather than a population estimate or backend ranking. It also drives the pooled differences: DeepSeek leads by seven verdicts and eight evidence validations, whereas GPT-5.5 leads across self-generalization and Pool~A by one verdict and three validations. We therefore interpret the tiers separately.
Specifically, Pool~B most clearly separates \(V\), \(I\), and \(E\): both backends identify all 37 targets, but only 26 GPT-5.5 reviews and 34 DeepSeek reviews satisfy evidence validation, while \(I>V\) shows that some target-identifying reviews have a non-blocking disposition. Pool~A shows the converse mismatch: both backends block four cases but identify the target in only three, so at least one block per backend is off-target. On the 50 paired benign controls, GPT-5.5 blocks five and DeepSeek blocks four.
\begin{table}[t]
    \centering
    \footnotesize
    \setlength{\tabcolsep}{2.5pt}
    \caption{Complete-pipeline Run~1 results on 63 held-out malicious cases and 50 paired benign controls. \(V/I/E\) are marginal counts, \(A=V\land I\land E\), and FP denotes a benign control blocked. GPT denotes GPT-5.5 and DS DeepSeek v4-pro. The pooled row is descriptive.}
    \label{tab:complete}
    \begin{tabular}{@{}llrrrrr@{}}
        \toprule
        Tier & Backend & \(V\) & \(I\) & \(E\) & \(A\) & FP \\
        \midrule
        Self-gen. & GPT
            & 13/19 & 14/19 & 12/19 & 12/19 & 0/6 \\
        & DS
            & 12/19 & 12/19 & 10/19 & 9/19 & 0/6 \\
        \midrule
        Pool A & GPT
            & 4/7 & 3/7 & 3/7 & 3/7 & 2/7 \\
        & DS
            & 4/7 & 3/7 & 2/7 & 2/7 & 1/7 \\
        \midrule
        Pool B & GPT
            & 26/37 & 37/37 & 26/37 & 26/37 & 3/37 \\
        & DS
            & 33/37 & 37/37 & 34/37 & 33/37 & 3/37 \\
        \midrule
        All & GPT
            & 43/63 & 54/63 & 41/63 & 41/63 & 5/50 \\
        & DS
            & 49/63 & 52/63 & 46/63 & 44/63 & 4/50 \\
        \bottomrule
    \end{tabular}
\end{table}

To check that this diagnosis advantage does not come from over-blocking, we also ran CodeRabbit on the six paired benign controls in the self-generalization tier: it blocked none (0/6), matching \sys/DeepSeek's 0/6 on the same controls. Both systems block 0/6 of these controls. This limited check provides no evidence that the observed identification difference is explained by greater blocking on the evaluated controls.

\noindent\textbf{Evidence-configuration ladder.}
We evaluate four evidence configurations on the 19 self-generalization cases.
E0 provides only the diff and disables the knowledge base. E1 adds touched
files, one-hop callers and callees, and the knowledge base. E2 provides the full structurally collected evidence and Stage~4 evidence closure, but disables the knowledge base. E3 is the complete \sys pipeline. Because several inputs
change between configurations, this is a configuration ladder rather than a
token-matched factorial ablation; we therefore interpret differences at the
configuration level.
\begin{table*}[t]
    \centering
    \small
    \caption{Performance across four evidence configurations on the 19 self-generalization cases.}
    \label{tab:ablation}
    \begin{tabular}{llrrr|rrr}
        \toprule
        & & \multicolumn{3}{c}{GPT-5.5}
          & \multicolumn{3}{c}{DeepSeek} \\
        \cmidrule(lr){3-5}\cmidrule(lr){6-8}
        Mode & Evidence & \(V\) & \(I\) & \(E\)
                       & \(V\) & \(I\) & \(E\) \\
        \midrule
        E0 & Diff
            & 9 & 13 & 6
            & 10 & 10 & 6 \\
        E1 & Touched + 1-hop + KB
            & 9 & 14 & 11
            & 12 & 11 & 10 \\
        E2 & Full evidence, no KB
            & 12 & 13 & 11
            & 11 & 11 & 10 \\
        E3 & Full \sys
            & 13 & 14 & 12
            & 12 & 12 & 10 \\
        \bottomrule
    \end{tabular}
\end{table*}
From E0 to E3, evidence validation rises from 6 to 12 cases for GPT-5.5 and from 6 to 10 for DeepSeek, whereas identification rises only from 13 to 14 and from 10 to 12, respectively. The observed gains therefore concentrate in validation, but the configurations change several inputs and do not isolate any component. E1 and E2 obtain the same validation counts despite differing in both evidence scope and knowledge-base availability; this comparison therefore does not identify the effect of either factor.

We answer RQ2 at the configuration level: relative to E0, the complete E3 configuration achieves higher evidence-validation counts for both backends, while identification changes more modestly. The observed difference is therefore concentrated in validation. However, because the ladder simultaneously changes evidence scope, stage separation, typed retrieval, evidence closure, and knowledge-base guidance, it does not identify which component causes the difference.

We additionally evaluate E0 and E3 on twelve held-out absence-type cases,
comprising all seven Pool~A cases and all five absence-type cases from the self-generalization tier. Under E0, GPT-5.5 blocks 1/12 and
DeepSeek 0/12; the sole block is off-target. Under E3, they block 6/12 and
5/12. For each backend, five cases change from a non-blocking verdict under E0 to \textsc{BLOCK} under E3, and none change in the reverse direction (exact McNemar, one-sided \(p=0.031\)). We treat this as a targeted
defect-class analysis rather than a population estimate; in particular, the
subset contains only one L2b case, which both backends miss.

\noindent\textbf{Does unrestricted repository access solve the problem?}
As a diagnostic probe of whether unrestricted repository search is sufficient in these cases, we run two independent commercial Codex-agent reviews on each of the seven malicious inputs. The agent receives the repository and diff but no \bench rubric, knowledge-base entry, later repository history, or hint about relevant peer code. It may search the full repository, invoke tools, and iterate without the hop limits imposed by \sys.
Across two runs on each of four L1 cases, six of eight runs return a blocking verdict, and the target vulnerability is identified for three of the four cases. Across two runs on each of three L2b cases, four of six runs return a blocking verdict, yet none of the delivered reviews identify the target vulnerability. Among six benign inputs, four of which are paired controls, one is blocked in both runs. Complete case-level results appear in Appendix~\ref{app:agent}.

The failure pattern is diagnostic despite the small, deliberately selected sample. Four of the six L2b runs issue a blocking verdict without identifying the target vulnerability, so the dominant observed failure is an off-target diagnosis. Verdicts also differ between the two runs for two of the three L2b inputs, whereas all four L1 inputs receive the same verdict in both runs. Thus, unrestricted repository search does not reliably yield a diagnosis that connects the changed code to the role-equivalent evidence required by these L2b cases. One of the three L2b inputs is the development-tier running example and contributes to no benchmark performance rate; we therefore treat this probe as qualitative evidence but not a system comparison with \sys.

\noindent\textbf{Cost, replayability, and stability.}
Normal \sys execution activates three mandatory model stages and, conditionally, the evidence-validation stage; bounded evidence closure and API or structured-output recovery may add calls. Across the paired external pools, no run exceeds 131k recorded tokens, and median end-to-end latency ranges from 79 to 167 seconds. At fixed Stage~0/Stage~1 inputs, Stage~2 is deterministic, reproducing identical normalized retrieval payloads across 168 replays.
However, end-to-end reviews remain nondeterministic. Across paired runs of 44 external-pool PRs, \(V\) changes in up to 25.0\% of cases, \(I\) in 6.8\%, and \(E/A\) in 15.9\%. Aggregate Run~1\(\rightarrow\)Run~2 \(I/E\) counts are \(40/29\rightarrow40/29\) for GPT-5.5 and \(40/36\rightarrow37/29\) for DeepSeek. These marginals complement the paired wobble counts because equal totals can mask case-level changes; Appendix~\ref{app:stability} reports the full breakdown.

\section{Real-World Vulnerability Discovery}
\label{sec:discovery}

This section answers RQ3: \emph{Can \sys uncover real-world vulnerabilities?} We apply \sys to real repository histories, manually validate the resulting findings, and then evaluate independent reviewers on the same fixed vulnerable PRs.
\sys uncovers twelve previously undisclosed, PoC-backed vulnerabilities across five widely used projects and five programming
languages. Nine are absence-type, and three are present-type. Their
evidence-location labels span all four levels defined in
Section~\ref{sec:bench_properties}: seven are L0, one is L1, three are L2a, and one is L2b. 

\subsection{Discovery Methodology and Findings}
\label{sec:discovery_method}

We use \sys as a security-review gate over a candidate stream derived from
repository history. The purpose is not to exhaustively scan every repository
revision, but to focus on expensive evidence-grounded review and human validation
on a manageable shortlist. The funnel is:

\begin{center}
\small
$\approx890$ mined $\rightarrow \approx80$ reviewed $\rightarrow 19$ blocked by \sys
$\rightarrow 14$ source-confirmed $\rightarrow 12$ PoC-backed
\end{center}

These twelve findings establish that \sys surfaces real, previously undisclosed vulnerabilities in production code. Because the candidate stream is filtered to concentrate manual validation effort, the study demonstrates discovery capability without estimating recall or expected yield on arbitrary PRs. Appendix~\ref{app:discovery} details the filtering and admission procedure.

All five repositories are absent from the \sys knowledge base. For every
admitted finding, we freeze the vulnerable revision, target vulnerability,
mechanism-level rubric, defect class, and evidence location before conducting the cross-system comparison. Verification is anchored in reproducible evidence, including repository history at pinned revisions, exhaustive source searches, and PoC reproduction.

\noindent\textbf{Findings.}
Table~\ref{tab:discovery} summarizes the twelve vulnerabilities. They span
authorization failures, SSRF, missing security enforcement, incorrect
matching logic, state-consistency errors, and insecure transport
configuration. Each finding is backed by a manually verified reproduction;
the table reports whether the executed PoC uses a lab setup, a running
instance, or a two-host experiment. CVSS values are
author-assigned rather than vendor or CNA scores. Due to the page limit, we select three representative cases to illustrate the practical and diagnostic roles of the discovery set.

\begin{table*}[t]
    \centering
    \small
    \caption{Previously undisclosed vulnerabilities surfaced by \sys.
    Project families are anonymized where required by coordinated disclosure.
    Class denotes absence-type (A) or present-type (P); Ev. denotes evidence
    location. All PoCs were executed; PoC reports the validation setting:
    Lab, Instance, or Two-host. CVSS values are author-assigned.}
    \label{tab:discovery}
    \begin{tabular}{r l l c c c c}
        \toprule
        \# & Project family (lang.) & Vulnerability mechanism (CWE)
        & Class & Ev. & PoC & CVSS \\
        \midrule

        1 & Payment integration (PHP)
          & Unverified caller of shared capture sink (1173/754)
          & A & L2a & Lab & 6.5 \\

        2 & Workflow platform (Rust)
          & Missing operator authorization on peer modules (862)
          & A & L2b & Instance & 7.6--8.5 \\

        3 & Automation platform (TS)
          & Allowed-domain guard not applied at model-search sink (693)
          & A & L2a & Instance & 7.1 \\

        4 & LLM gateway (Python)
          & Redirecting SSRF sink omitted from URL guard (918)
          & A & L1 & Lab & 7.7 \\

        5 & Automation platform (TS)
          & Resume IDOR via client-supplied run identifier (639)
          & A & L2a & Instance & 6.4 \\

        6 & LLM gateway (Python)
          & SSRF guard covers only two of eight methods (918/200)
          & A & L0 & Lab & 7.7 \\

        7 & CMS (C\#)
          & Notification BOLA enabling host takeover (639/863)
          & A & L0 & Instance & 8.8 \\

        8 & Payment integration (PHP)
          & Order BOLA without object-level authorization (639)
          & A & L0 & Two-host & 9.4 \\

        9 & Payment integration (PHP)
          & Wildcard payment over-match (155/863)
          & P & L0 & Two-host & 8.6 \\

        10 & LLM gateway (Python)
           & Model-budget TOCTOU (362/770)
           & P & L0 & Lab & 4.3 \\

        11 & LLM gateway (Python)
           & Unbounded in-flight budget spend (770/799)
           & P & L0 & Lab & 4.3 \\

        12 & Payment integration (PHP)
           & SFTP without host-key verification (322)
           & A & L0 & Lab & 6.8 \\

        \bottomrule
    \end{tabular}
\end{table*}

\begin{itemize}
\itemsep0em
    \item \emph{Cross-project resume IDOR (\#5).}
    A new resume endpoint accepts a body-supplied \texttt{runId}. Although the
    route has project-scoped authorization, that check covers URL parameters
    only, while the unchanged checkpoint store resolves \texttt{runId} without
    an agent or project predicate. Given a valid victim-issued UUID, an executed
    PoC confirmed that a user in one project could resume another project's
    checkpoint, exposing its serialized conversation and pending tool-call
    arguments and allowing the attacker to supply the human-in-the-loop
    decision. Locating the unscoped lookup requires following the changed
    endpoint into an unchanged component, making this an L2a case. CodeRabbit
    blocks the PR but reports an unrelated client-side secret-rendering issue
    rather than the missing ownership binding.

    \item \emph{Notification BOLA leading to host takeover (\#7).}
    A notification-update handler authorizes a body-supplied recipient
    identifier, which an attacker can set to their own user ID, but then reloads
    and returns a different stored notification selected by an unchecked
    identifier. Sequential notification IDs and an automatically assigned
    registered-user role allow a self-registered account to enumerate other
    users' notification bodies, including password-reset, e-mail-verification,
    and second-factor tokens. The executed PoC used a disclosed host
    password-reset token to reset the password and authenticate with Host and
    administrator privileges.

    \item \emph{Cross-customer order BOLA (\#8).}
    Public shop endpoints resolve a client-supplied, sequential database order
    ID without binding the selected order to the caller or using the existing
    unguessable order token. The handler can replace the victim order's customer
    and shipping and billing addresses, advance its checkout state, and persist
    those changes. An executed two-host PoC confirmed that an unauthenticated
    attacker could mutate another customer's in-flight order, enabling
    guest-cart takeover, payment manipulation, and diversion of goods. The
    finding was subsequently confirmed by the vendor.
\end{itemize}

\subsection{The VD Gap on Real-World Vulnerabilities}
\label{sec:discovery_vd}
The twelve validated vulnerabilities form the Discovery tier in
Table~\ref{tab:bench} and test whether the VD gap extends beyond curated
benchmark constructions to real-world vulnerabilities. On this tier,
independently run \sys/DeepSeek and CodeRabbit both issue blocking verdicts on
10/12 cases. Their diagnoses diverge sharply: \sys/DeepSeek obtains
\(V=10/12\), \(I=10/12\), and \(E=10/12\), whereas CodeRabbit obtains
\(V=10/12\), \(I=4/12\), and \(E=4/12\). CodeRabbit's \(I\) and \(E\) scores
coincide because every target-identifying delivered comment also satisfies the
evidence-validation requirements. Thus, identical verdict totals yield
\(A=10/12\) for \sys/DeepSeek and \(A=4/12\) for CodeRabbit. These tier-level
scores are included in the 31-case aggregates reported in
Tables~\ref{tab:defect} and~\ref{tab:common-attribution}; the non-discovery
evaluation in Table~\ref{tab:complete} excludes this tier by design.
Accordingly, six of CodeRabbit's ten blocks are triggered by findings that do not
identify the vulnerability that makes the PR unsafe.
In each of those cases, the reviewer
raises a security alarm, but repairing the reported issue does not necessarily
remove the validated target vulnerability. The VD gap can therefore make an
automated reviewer appear successful while leaving the security-critical
defect unidentified. Combined with the controlled results of
Section~\ref{sec:vd_results}, the discovery study shows that this failure mode
is not merely a consequence of synthetic or advisory-derived benchmark
construction.

\noindent\textbf{Evidence beyond the touched files.}
The discovered vulnerabilities also illustrate why repository evidence matters
for diagnosis. Four of the twelve are classified as L2a or L2b and therefore require validation evidence outside the touched files. They correspond to different repository relationships rather
than one recurring vulnerability template.
One finding requires following callers of a shared payment operation to reveal
an unverified third caller. Another requires comparing role-equivalent modules
whose authorization requirements are not connected by a direct call,
definition, or import edge. A third requires determining which model-search
paths never consume an allowed-domain guard. The resume IDOR requires following a dependency into an
unchanged checkpoint store and contrasting its owner-free lookup with an
ownership-bound peer.
These cases respectively exercise the caller, sibling, consumer, and guard
relationships targeted by \sys's typed evidence acquisition. At the same time,
seven of the twelve findings are L0 and another is L1, showing that the
discovery set is not restricted to vulnerabilities requiring non-local
evidence. Rather, the findings demonstrate that a practical PR reviewer must
handle both vulnerabilities visible near the change and those whose diagnosis
depends on repository context outside it.

\noindent\textbf{Validation and controls.}
Every reported vulnerability is manually validated before inclusion in the
challenge set. Appendix~\ref{app:cves} reports the coordinated-disclosure
status of each case. As an additional check against indiscriminate blocking, the complete configurations are also evaluated
on the 50 paired complete-fix controls from \bench. In the reported run,
GPT-5.5 and DeepSeek produce five and four false positives, respectively,
passing 45/50 and 46/50 controls.

Overall, the discovery study answers RQ3 affirmatively: the
Verdict--Diagnosis gap occurs on genuine, previously undisclosed
vulnerabilities. A reviewer can correctly block a vulnerable PR while failing
to identify the vulnerability that actually requires remediation.

\section{Discussion}
\label{sec:discussion}

\noindent\textbf{Verdict and diagnosis capture different aspects of review quality.}
On the 31-case common-coverage challenge set, CodeRabbit blocks 24 PRs and
\sys/DeepSeek blocks 22, yet \sys/DeepSeek identifies 22 target vulnerabilities
compared with CodeRabbit's 16. The difference is concentrated in absence-type
defects: both systems block 9/14 cases, while \sys/DeepSeek identifies 9 targets
and CodeRabbit identifies 3. Output-level attribution preserves this
distinction: \sys/DeepSeek produces 19 attributable blocks and CodeRabbit 16.
These results motivate evaluating PR security review beyond the verdict
and treating vulnerability identification and evidence validation as separate
outcomes.

\noindent\textbf{Diagnosis becomes harder when security evidence is implicit or non-local.}
The VD gap is concentrated in absence-type defects and L2a/L2b cases, where
validating the target vulnerability requires evidence outside the touched
files. These cases require the reviewer to infer missing enforcement or connect
the change to security-relevant repository context that is not explicit in the
diff. The same pattern appears across the held-out evaluation, suggesting that
these properties are helpful for characterizing PR-review
difficulty.

\noindent\textbf{Validation is stronger in evidence-rich configurations.}
Across the configuration ladder, moving from diff-only review to the full
pipeline changes vulnerability identification modestly while evidence
validation improves substantially for both backends. Providing more repository
context does not consistently improve performance, a pattern consistent with
the value of selecting relevant evidence rather than maximizing context volume.
Because E0--E3 change several factors simultaneously, however, these results do
not isolate the causal contribution of the knowledge base, which a token-matched on/off ablation is left to establish.

\noindent\textbf{Repository access alone is insufficient.}
The unrestricted-agent probe further highlights the role of evidence
selection. The agent identifies 3/4 target vulnerabilities among the L0/L1
cases, but none of the three L2a/L2b vulnerabilities, despite blocking all
three at least once. Together with the configuration-level retrieval results,
this suggests that effective PR review requires not only access to repository
context, but also a mechanism for selecting the callers, consumers, guards, or
peers relevant to the security question.

\noindent\textbf{Why these vulnerabilities matter.}
The discovery set is not a collection of synthetic or low-severity artifacts. All twelve are previously undisclosed, PoC-backed defects in production repositories across five widely used projects and five languages, with impacts including cross-project data access, object-level authorization bypass, and host-account takeover (Table 7). All were reported through coordinated disclosure, and several have been vendor-confirmed (Table 12). Their relevance to this paper is not their individual severity but where the VD gap appears: a widely deployed commercial reviewer correctly blocks these PRs yet, in six of ten cases, reports a finding that does not identify the defect a maintainer must fix. The failure mode therefore reaches genuine, high-impact vulnerabilities in code that real projects ship and maintain, not only curated benchmark constructions.

\noindent\textbf{Scope.}
The deployed-reviewer comparison covers 31 common-coverage challenge cases and
stratifies them by defect class and evidence location. These properties
characterize what a correct validation must establish and how far beyond the
diff a reviewer must look to obtain the required repository facts. The
production study separately establishes the practical relevance of attributable
review through twelve PoC-backed vulnerabilities. Since those twelve cases
were surfaced by the originating \sys/GPT-5.5 configuration, they test whether
the VD gap appears in fixed real-world cases; they do not provide an unselected
estimate of general \sys-versus-CodeRabbit superiority.

\section{Related Work}
\label{sec:related}

\noindent\textbf{Vulnerability-review benchmarks.}
Most vulnerability datasets assign a binary label to a function or
commit~\cite{bigvul}; subsequent work shows that noisy labels and unrealistic
class balance can inflate apparent performance~\cite{primevul}. The closest
review benchmark evaluates LLMs and tool-using agents on paired vulnerable and
benign commits~\cite{jitvul}. It measures detection and observes off-target
speculation, but does not make the reported mechanism an independently scored
outcome. SecLLMHolmes similarly finds that a correct vulnerability verdict need
not be supported by faithful reasoning~\cite{secllmholmes}. \bench brings this distinction to PR review: every malicious case fixes a target vulnerability and validation requirements and scores verdict, vulnerability identification, and evidence validation separately. Defect class and evidence location characterize the cases in which the \gap occurs.

\noindent\textbf{Code Security Analysis.}
Static analyses have long found omitted checks by comparing related code.
Chucky contrasts a function with semantic neighbors~\cite{chucky}; RoleCast
groups web-application code by role to infer missing enforcement without a
known check~\cite{rolecast}; and APISan infers intended API checks from sibling
usages~\cite{apisan}. These systems establish the value of peer comparison. Our
setting asks a different question: when reviewing one PR, can a reviewer locate
the relevant reference and explain the missing guard, binding, or state update?
The reference may be in the touched file, dependency-reachable code, or a
role-related peer with no dependency edge to the change. \bench captures this
variation through absence-type defects and evidence location; \sys turns
repository knowledge into a bounded, role-typed peer query at review time.

\noindent\textbf{Agents and retrieval.}
ReAct, Self-RAG, and FLARE interleave generation with run-time decisions to
retrieve or act~\cite{react,selfrag,flare}. Software-engineering evaluations
likewise compare agentic scaffolds with fixed pipelines on issue resolution,
where tests provide an execution oracle~\cite{swebench,agentless}. \sys instead
operates on an adversarial change with no such oracle. It collects code through
deterministic, non-executing tools, uses build-time knowledge for type retrieval,
and separates vulnerability identification from evidence validation. The
evaluation thus concerns the attribution of a security review, rather
than a general ranking of agent architectures.

\noindent\textbf{Adjacent security tasks.}
CyberGym and ExploitGym evaluate whether agents can reproduce or weaponize a
vulnerability supplied as input~\cite{cybergym,exploitgym}; IssueTrojanBench
studies coding agents that receive malicious issue instructions~\cite{issuetrojan};
and security-hardened generation acts on the producing side of the
workflow~\cite{sven}. Our attacker instead submits a malicious PR whose
vulnerability is unknown to the reviewer. The reviewer remains read-only and
must decide whether to block the change and provide a diagnosis that identifies
the threat the maintainer must fix.

\section{Conclusion}
\label{sec:conclusion}

Automated security review should be judged not only by whether it blocks a
malicious PR, but by whether it identifies the vulnerability that justifies the
block. We introduced \bench to measure this verdict--diagnosis gap through
separate verdict, vulnerability-identification, and evidence-validation scores,
and \sys to test whether staged, typed evidence retrieval can narrow it. Our
results show that verdict-level performance can substantially overstate review
quality. On the 31 common-coverage challenge cases, \sys and CodeRabbit
produce similar blocking totals, yet \sys identifies more target
vulnerabilities. The difference is largest for absence-type defects, where it
identifies three times as many. The same pattern appears on the twelve validated
production vulnerabilities, all previously undisclosed and PoC-backed:
independently run \sys and CodeRabbit produce identical blocking totals,
but \sys yields 2.5$\times$ as many attributable blocks. Together,
these findings show why verdict-only evaluation is insufficient and motivate
review systems that tell maintainers what must be fixed and where the supporting
evidence lies.

\appendix
\cleardoublepage
\section{Ethical Considerations}
\label{sec:ethics}

This work analyzed public source code and researcher-controlled local or test
instances. We accessed no user data and did not probe production services.
Previously undisclosed findings are handled through coordinated disclosure:
reports provide maintainers or an appropriate coordinator with the mechanism,
affected revision, and a proof of concept where safe. Unfixed or
unacknowledged findings are anonymized, and reproduction detail is withheld.
Project names, identifiers, and credit are restored only when disclosure status
permits. Appendix~\ref{app:cves} records the per-finding evidence and disclosure
boundary; the final submission will freeze that table against the disclosure
ledger.

The study also sends public repository content to hosted models. Repositories
containing confidential code are outside our experiment. A deployment can use
the open-weights backend on premises; the security boundary we claim is about
reviewed-input execution and store writes, not provider confidentiality. We
judge the defensive benefit of documenting the measurable failure mode to
outweigh residual risk, while limiting operational exploit detail until fixes
are available.

\bibliographystyle{plain}
\bibliography{bibliography}

\section{Reproducibility Records}
\label{app:artifact-boundary}

The artifact maps every reported aggregate to a frozen runner configuration,
repository revision, result manifest, and pre-committed rubric. It also records
the ordered patch stack, environment toggles, sealed-worktree metadata, and
stage-level cost and output logs. Section~\ref{app:open-science} describes the
release boundary for unfixed vulnerabilities.

\section{External Advisory Pool Construction}
\label{app:external-construction}

The external pools begin with GitHub-reviewed advisories published in 2025.
We resolve each advisory's fix commit and parent revision following the
CVEfixes collection methodology~\cite{cvefixes}. Candidate admission is fixed
before model execution; duplicate mechanisms, unreachable residuals, diffs
above 40~KB, overlap between pools, and defense-in-depth changes are excluded
under recorded criteria. Table~\ref{tab:external-funnel} gives the resulting
selection accounting.
\begin{table}[t]
\centering
\caption{External advisory harvest and Pool~A/Pool~B construction gates.}
\label{tab:external-funnel}
\small
\begin{tabular}{@{}llr@{}}
\toprule
Stage & Filter & Count \\
\midrule
-- & 2025 GitHub-reviewed advisories & 3,688 \\
N0 & Linkable GitHub fix commit & 2,231 \\
N1 & Web ecosystem, authorization CWE & 237 \\
N1$'$ & After exfiltration blacklist & 227 \\
N2 & Fix and parent resolvable & 213 \\
NA & Omittable enforcement hunk & 11 \\
NB & Single-site revertible guard & 50 \\
\midrule
Admitted A/B & Fixed manual quality gate & 7 / 37 \\
\bottomrule
\end{tabular}
\end{table}

\section{External Pool A in Full}
\label{app:poolA}

Table~\ref{tab:poolA-full} lists the complete Pool~A tier. Each row is anchored
to a disclosed access-control advisory fix. The malicious state retains the
production hunks but omits the listed enforcement hunk or file; the benign twin
is the complete fix. Identification credit requires the omitted enforcement
mechanism rather than a generic or neighboring authorization finding.
``Sites'' counts the relevant enforcement or binding sites in the real fix.

\begin{table*}[!t]
\centering
\caption{The seven evaluated External Pool~A pairs and one parked candidate.}
\label{tab:poolA-full}
\footnotesize
\setlength{\tabcolsep}{3pt}
\renewcommand{\arraystretch}{1.08}
\begin{tabular}{@{}rL{0.19\textwidth}L{0.12\textwidth}rrL{0.46\textwidth}@{}}
\toprule
\# & Repository (lang) & Advisory & CWE & Sites & Site left unguarded / omitted deny \\
\midrule
1 & \path{jenkinsci/blazemeter-plugin} (Java) & CVE-2025-13472 & 862 & 3 & \texttt{BlazeMeterPerformanceBuilderDescriptor.java} --- \texttt{item.checkPermission(Item.READ)} \\
2 & \texttt{jenkinsci/}\newline\texttt{global-build-stats-}\newline\texttt{plugin} (Java) & CVE-2025-58459 & 284 & 2 & \texttt{GlobalBuildStatsPlugin.java} --- \texttt{checkPermission(getRequiredPermission())} \\
3 & \path{jenkinsci/opentelemetry-plugin} (Java) & CVE-2025-58460 & 862 & 3 & \texttt{ElasticLogsBackendWithJenkinsVisualization.java} --- \texttt{if (!isAuthorized())} \\
4 & \path{navidrome/navidrome} (Go) & CVE-2025-48948 & 863 & 4 & \path{persistence/transcoding_repository.go} --- \texttt{!isAdmin(ctx)} $\rightarrow$ \texttt{ErrPermissionDenied} \\
5 & \path{smallstep/certificates} (Go) & CVE-2025-44005 & 306 & 3 & \path{authority/tls.go} --- \texttt{revokeOpts.Serial != claims.Subject} $\rightarrow$ \texttt{errs.Forbidden(...)} \\
6 & \path{usememos/memos} (Go) & CVE-2025-65795 & 284 & 6 & \path{server/router/api/v1/memo_attachment_service.go} --- \texttt{CreatorID != user.ID} $\rightarrow$ \texttt{PermissionDenied} \\
7 & \path{TYPO3-CMS/backend} (PHP) & CVE-2025-59017 & 862 & 1 & \path{Classes/Middleware/BackendModuleValidator.php} --- the \texttt{inheritAccessFromModule} enforcement branch \\
\midrule
-- & \texttt{valtimo-platform/}\newline\texttt{valtimo-backend-}\newline\texttt{libraries} (Java) & CVE-2025-48881 & 863 & 8 & \emph{parked}: \texttt{ObjectenApiClient.kt} --- \texttt{requirePermission(...)}; independent reachability of the omitted site could not be established \\
\bottomrule
\end{tabular}
\end{table*}

Every omitted site lies outside the changed lines. Five require unchanged code
in a touched file (\texttt{BlazeMeter}, \texttt{global\_build\_stats},
\texttt{OpenTelemetry}, \texttt{Navidrome}, and \texttt{memos}); two require
evidence outside the touched files (\texttt{smallstep} and \texttt{TYPO3}). All
seven are therefore non-diff-local, while the latter two additionally test
cross-file evidence acquisition.

\paragraph{The tier's construction contract.}
Every pair keeps the real fix commit's production hunks and strips tests from both
sides. The omitted enforcement lies outside the maliciously changed lines, and its
implementation does not appear in the remaining hunks, although declarations that
depend on it may remain as evidence. If a whole file is omitted,
its pre-existing code must remain byte-for-byte at the base revision so that the
construction removes only new enforcement rather than weakening an existing check.
The artifact rubric records the exact omitted decision for each pair.

\section{Case-Level Evidence Locations}
\label{app:evidence-loc}

Defect class (absence/present) and evidence location (L0/L1/L2a/L2b) capture
different case properties. Table~\ref{tab:app-sg-class} reports both for the
full self-generalization tier, making the twelve-case absence analysis of
Section~\ref{sec:understanding} auditable at case level.
Class is abbreviated A/P. ``Mode'' is the typed second-pass directive recorded
by the case rubric; --- denotes no second retrieval pass.

\begin{table*}[!t]
\centering
\caption{Case-level defect class, evidence location, and retrieval mode for the
self-generalization tier.}
\label{tab:app-sg-class}
\footnotesize
\setlength{\tabcolsep}{3pt}
\renewcommand{\arraystretch}{1.04}
\begin{tabular}{@{}llll@{\qquad}llll@{}}
\toprule
Case & Cl. & Loc. & Mode & Case & Cl. & Loc. & Mode \\
\midrule
\texttt{directus\_53889} & A & L0 & --- &
\texttt{directus\_GHSA-3573} & P & L0 & --- \\
\texttt{directus\_cff8} & A & L0 & --- &
\texttt{dify\_9906} & P & L1 & --- \\
\texttt{windmill\_26964} & A & L1 & \textsc{consumer-sink} &
\texttt{dify\_weak\_random} & P & L0 & --- \\
\texttt{apisix\_24112} & A & L2a & --- &
\texttt{n8n\_webhook\_ip\_substring} & P & L0 & --- \\
\texttt{n8n\_expr\_isolate} & A & \textbf{L2b} & \textsc{sibling-endpoint} &
\texttt{windmill\_23696} & P & L0 & --- \\
\midrule
\texttt{apisix\_29266} & P & L0 & --- &
\texttt{windmill\_29059} & P & L1 & \textsc{caller-source} \\
\texttt{apisix\_46647} & P & L0 & --- &
\texttt{windmill\_p5cj} & P & L0 & --- \\
\texttt{apisix\_62232} & P & L0 & --- &
\texttt{n8n\_loadoptions} & P & L2a & \textsc{consumer-sink} \\
\texttt{directus\_39701} & P & L0 & --- &
\texttt{apisix\_admin\_default} & P & L0 & --- \\
\texttt{directus\_55746} & P & L0 & --- & & & & \\
\bottomrule
\end{tabular}

\vspace{3pt}
\begin{minipage}{0.96\textwidth}
\footnotesize
The tier contains 19 cases over five repositories: Directus~5, Dify~2,
APISIX~5, Windmill~4, and n8n~3. The five A rows are absence-type, and the
fourteen P rows are present-type. \texttt{apisix\_admin\_default} is an
nginx-template case and is therefore not part of the four-case Lua slice in
Table~\ref{tab:lua}.
\end{minipage}
\end{table*}

\section{Sibling-Shaped Cases, With Their Evidence}
\label{app:sibling}

The reachability argument is carried by named cases rather than by a rate, so we therefore report the qualifying cases together with near misses and the criterion excluding each one. For each, we give the reviewed revision, the touched-file set, the peer
site's location, and the outcome of the three dependency queries run on the base
revision: a case qualifies only if no query returns a set that isolates the peer.
``Direction'' distinguishes a peer that carries the security anchor
(contrast) from a peer at which the anchor is absent (omission). ``Isolating
edge'' asks whether a call, definition, or import query from a touched file
returns a set that singles out the peer.
Of the three qualifying rows, one is a development case, and one never blocks;
\texttt{windmill\_22683} is the qualifying case in a scorable tier. We therefore
treat sibling-only reachability as a case-level existence result rather than a
rate.
\FloatBarrier

\begin{table*}[!t]
\centering
\caption{Sibling-shaped evidence cases and boundary exclusions.}
\label{tab:app-sibling}
\label{tab:app-sibling-excluded}
\footnotesize
\setlength{\tabcolsep}{3pt}
\renewcommand{\arraystretch}{1.08}
\begin{tabular}{@{}L{0.13\textwidth}L{0.065\textwidth}L{0.08\textwidth}rL{0.50\textwidth}L{0.12\textwidth}@{}}
\toprule
Case & Tier & Direction & $|T|$ & Peer site and non-isolation evidence & Outcome \\
\midrule
\texttt{authentik\_9076} & constr. & contrast & 4 &
\path{providers/oauth2/views/authorize.py:332},
\texttt{AuthorizationFlowInitView} (\texttt{PolicyAccessView}). The four
touched files contain no reference to either view or to \texttt{authorize}. &
n/a (development tier) \\
\addlinespace
\texttt{windmill\_22683} & disc. & omission & 4 &
eight \texttt{create\_*}/\texttt{update\_*} handlers in \texttt{folders.rs},
\texttt{groups.rs}, \texttt{resources.rs}, \texttt{schedule.rs},
\texttt{triggers/handler.rs}, and \texttt{variables.rs}. Caller queries stay
inside the touched files; the shared \texttt{check\_scopes} has 92 call sites
against a peer set of eight. &
\checkmark{} (both backends) \\
\addlinespace
\texttt{n8n\_expr\_isolate} & SG & omission & 2 &
unwired call sites in \path{packages/core/src/execution-engine/}. The touched
service imports no execution-engine symbol. &
\textbf{never blocks} \\
\bottomrule
\end{tabular}

\vspace{6pt}
{\footnotesize\emph{Boundary cases excluded from the sibling-shaped set; each
row records the criterion that fails.}}\par\vspace{2pt}
\footnotesize
\setlength{\tabcolsep}{3pt}
\renewcommand{\arraystretch}{1.08}
\begin{tabular}{@{}L{0.15\textwidth}L{0.07\textwidth}L{0.12\textwidth}L{0.62\textwidth}@{}}
\toprule
Case & Tier & Failed check & Reproducible reason \\
\midrule
\texttt{n8n\_searchmodels} & disc. & isolating edge &
The touched node imports \texttt{searchModels} directly from
\path{methods/loadModels.ts}. \\
\texttt{authentik\_12900} & constr. & isolating edge &
Touched \path{core/models.py:345} imports \texttt{UserSerializer} from
\path{core/api/users.py:115}. \\
\texttt{dify\_9906} & SG & outside $T$ &
The peer in \path{controllers/console/files/__init__.py} is inside the touched
file set. \\
\texttt{zitadel\_7963} & constr. & outside $T$ &
\texttt{DeleteSession} is in the same touched file, 25 lines from the change. \\
\texttt{paypal\_capture} & disc. & isolating edge &
A reverse call-graph query returns exactly the three callers of the shared
payment sink. \\
\texttt{litellm\_url\_guard} & disc. & outside $T$ &
\texttt{get\_base\_url} and \texttt{create\_tool\_function} are in the same
touched file as the guarded specification fetch. \\
\texttt{n8n\_agents\_resume} & disc. & outside $T$ &
A walk reaches the checkpoint store but cannot isolate an absent owner key;
the contrast at \texttt{agents.controller.ts:578--590} is inside the diff. \\
\bottomrule
\end{tabular}
\end{table*}

\section{Discovery-Set Construction and Disclosure Boundary}
\label{app:discovery}
\label{app:cves}

\noindent\textbf{Construction.}
Deterministic fix-history mining produces approximately 890 candidate PR
states. An LLM subagent applies a grouped shape filter, processing candidates
in batches and recording group-level exclusion reasons. This stage evaluates
whether a candidate has a structure worth further investigation rather than
deciding whether a vulnerability exists. It removes candidates when
(i) the issue and decisive evidence are entirely diff-local;
(ii) no earlier or later fix, protected peer, guarded path, or other useful
comparator can be found; or
(iii) there is no indication that the relevant path is reachable from
untrusted input. The filter reduces the stream to approximately 80 candidates
for full \sys/GPT-5.5 review.

Of these candidates, \sys blocks 19. Manual source confirmation followed by an
independently prompted refuter retains 14 source-confirmed findings. We attempt
to construct and execute PoCs for all 14; two for which PoC construction cannot
be completed are excluded. The final set contains 12 previously undisclosed
vulnerabilities from five production repositories absent from the \sys
knowledge base. Every admitted finding has a manually verified, executed PoC
and a complete attack chain. After admission, we freeze its repository revision
and mechanism-level rubric before running \sys/DeepSeek and CodeRabbit.

\noindent\textbf{Disclosure status.}
All twelve findings were reported before submission. Table~\ref{tab:disclosure}
records only the report and maintainer-response status available at the
disclosure-status freeze. We omit remediation status because it has not been
uniformly reverified. No CVE or GHSA identifier had been assigned at the
freeze, and operational reproduction details are withheld.

\begin{table}[t]
\centering
\small
\caption{Coordinated-disclosure status at the disclosure-status freeze. Case
numbers correspond to Table~\ref{tab:discovery}.}
\label{tab:disclosure}
\begin{tabular}{@{}cll@{}}
\toprule
Cases & Project family & Disclosure status \\
\midrule
1, 8, 9, 12
    & Payment integration
    & Reported; vendor-confirmed \\
2
    & Workflow platform
    & Reported; awaiting response \\
3, 5
    & Automation platform
    & Reported; vendor-confirmed \\
4, 6, 10, 11
    & LLM gateway
    & Reported; awaiting response \\
7
    & CMS
    & Reported; vendor-confirmed \\
\bottomrule
\end{tabular}
\end{table}

\section{Grading Details}
\label{app:grading}
A review naming an accepted mechanism but denying its security impact earns
identification but not evidence-validation credit. A neighboring real defect
earns neither score unless pre-listed as equivalent. CodeQL is graded from its
alert rule and dataflow path under the same content requirements.

\section{Implementation Parameters}
\label{app:implementation}
The isolated runner records the pipeline version, model/provider, temperature,
run index, repository revision, selected mechanism and knowledge-entry IDs,
structural counts, stage outputs, elapsed time, token count, and verdict. Stage~0
enumerates every path touched by the raw diff, while its primary per-touched-file
scans process the first 24 paths in diff order. The resulting context pack retains
at most 16 hunk excerpts, eight same-file callees, eight contract references, six
importer excerpts, three upstream producers, three contrast siblings, three
co-emitting siblings, two verified consumers, two callers, four cross-file
callees, 14 related-file excerpts, 24 peer-class records, and 24 symbol-reference
records. A hunk excerpt is not the diff hunk itself: it is a post-change
repository snippet expanded from an added-line location to the enclosing
function, or to a bounded window when no function is recognized. The prompt
materializer considers the first ten such excerpts under a 16,000-character
section budget and a 48,000-character context-pack budget; the raw diff is
supplied separately to Stages~1, 3, and 4. The 24-path scan frontier and
repository-search truncations emit uncertainty records. Context-pack and prompt
materialization limits are recorded by their fixed configuration rather than by
a per-run uncertainty flag.

\paragraph{Knowledge-base inventory.}
The frozen KB contains 21 mechanisms and 25 entries. Seventeen mechanisms
contain one entry and four contain two. Seven entries provide guidance to
Stage~3 and eighteen to Stage~4; sixteen use \texttt{single} retrieval and
nine use \texttt{two\_stage} retrieval.

Retrieval uses \texttt{BAAI/bge-small-en-v1.5} at revision
\texttt{5c38ec7c405ec4b44b94cc5a9bb96e735b38267a}. Coarse retrieval keeps top-5
at threshold 0.35; fine retrieval keeps top-3 at threshold 0.50 with
$0.8\,s_{coarse}+0.2\,s_{structural}$. Repository loopback scans at most 5,000 files,
returns 20 hits, and injects up to 12 compact evidence items. On the normal
control path, Stages~1, 3, and 5 call the model and Stage~4 is conditional.
Evidence closure lets Stage~4 request at most three named symbols per round and
repeat for at most three rounds, with at most eight fetched definitions in
total; per-stage API retry and JSON repair are separately capped and recorded.
The verdict mapping is given in \S\ref{sec:verdict_synthesis}.

\section{Commercial-Agent Probe}
\label{app:agent}
Two fresh agents receive only a neutral diff and repository at the reviewed
revision. The prompt does not mention this paper, siblings, omissions, expected
maliciousness, rubrics, prior results, or the knowledge base. Network access and
future history are unavailable; repository search and autonomous iteration are
unrestricted.

\begin{table}[htbp]
\centering
\caption{Commercial Codex probe, two blind runs. Target identification uses the
same rubric as all systems.}
\label{tab:abl-e2}
\small
\begin{tabular}{@{}llccc@{}}
\toprule
Case & Evid. & Run 1 & Run 2 & Target (I) \\
\midrule
BlazeMeter & L1 & block & block & \checkmark \\
global build stats & L1 & block & block & \checkmark \\
memos & L1 & block & block & \checkmark \\
Navidrome & L1 & approve & approve & -- \\
\midrule
authentik 9076 & L2b & approve & block & -- \\
windmill 22683 & L2b & block & block & -- \\
n8n expr isolate & L2b & block & approve & -- \\
\midrule
L1 subtotal & & \multicolumn{2}{c}{6/8 blocking runs} & 3/4 \\
L2b subtotal & & \multicolumn{2}{c}{4/6 blocking runs} & 0/3 \\
\bottomrule
\end{tabular}
\end{table}

These blind repetitions are a diagnostic probe and are not aggregated into the
main performance scores. Of six benign inputs, one blocks in both runs; only
four are genuine paired twins. Across all 26
runs, tool invocations range from 12 to 47 and files read from 5 to 23; the same
input varies by up to $2.2\times$ in invocation count. Wall time is available for
23 runs and spans 1m22s--10m08s. Exact token consumption is not exposed by the
product interface.

\section{Cost, Replay, and Run-to-Run Stability}
\label{app:stability}
\begin{table*}[t]
\centering
\caption{Measured \sys cost per run. GPT token counts are estimated;
DeepSeek counts are provider-reported. ``4-call'' denotes normal-path runs;
bounded closure or recovery may add calls.}
\label{tab:g4}
\small
\begin{tabular}{@{}lrrrrrr@{}}
\toprule
& 4-call & \multicolumn{3}{c}{Tokens} & \multicolumn{2}{c}{Elapsed (s)} \\
\cmidrule(lr){3-5}\cmidrule(lr){6-7}
Config & runs & Median & P90 & Max & Median & P90 \\
\midrule
GPT-5.5, Pool A & 57\% & 38.7k & 92.6k & 97.4k & 107 & 201 \\
GPT-5.5, Pool B & 73\% & 51.3k & 85.0k & 112.6k & 167 & 290 \\
DeepSeek, Pool A & 50\% & 40.6k & 103.1k & 103.6k & 79 & 154 \\
DeepSeek, Pool B & 68\% & 49.1k & 82.0k & 130.7k & 159 & 266 \\
\bottomrule
\end{tabular}
\end{table*}

\begin{table*}[t]
\centering
\caption{Run-to-run label stability on malicious PRs from the external pools.
An \(X\) wobble denotes a case-level difference in label \(X\) between the
predesignated Run~1 review and the independent rerun. Attribution is recomputed
for each run as \(A=V\land I\land E\).}
\label{tab:secprops}
\small
\begin{tabular}{@{}lrrrrr@{}}
\toprule
& & \multicolumn{4}{c}{Wobble} \\
\cmidrule(lr){3-6}
Config & $N$ & $V$ & $I$ & $E$ & $A$ \\
\midrule
GPT-5.5, Pool A & 7
    & 2/7 (28.6\%) & 2/7 (28.6\%) & 2/7 (28.6\%) & 2/7 (28.6\%) \\
GPT-5.5, Pool B & 37
    & 6/37 (16.2\%) & 0/37 (0\%) & 0/37 (0\%) & 1/37 (2.7\%) \\
GPT-5.5, A+B & 44
    & 8/44 (18.2\%) & 2/44 (4.5\%) & 2/44 (4.5\%) & 3/44 (6.8\%) \\
\midrule
DeepSeek, Pool A & 7
    & 3/7 (42.9\%) & 2/7 (28.6\%) & 1/7 (14.3\%) & 1/7 (14.3\%) \\
DeepSeek, Pool B & 37
    & 8/37 (21.6\%) & 1/37 (2.7\%) & 6/37 (16.2\%) & 6/37 (16.2\%) \\
DeepSeek, A+B & 44
    & 11/44 (25.0\%) & 3/44 (6.8\%) & 7/44 (15.9\%) & 7/44 (15.9\%) \\
\bottomrule
\end{tabular}
\end{table*}

Static counts include collected paths, code blocks, sibling records, and
call-graph records and are identical across paired runs. The stability table
binarizes \texttt{metadata.final\_verdict} after the deterministic guard as
\textsc{BLOCK} versus non-blocking; \textsc{COMMENT}-to-\textsc{APPROVE}
changes therefore do not count as V wobble. A divergent Stage~5 prose verdict
is recorded but does not replace this field. The \(I\) and \(E\) columns compare
the final adjudicated rubric labels for the two delivered reviews, and \(A\) is
recomputed separately for each run. The
Stage~2 fixed-input control reads the saved Stage~0 and
Stage~1 objects, diff path, and repository root from every Pool~A run JSON,
loads the same local ONNX encoder and corpus-locked index, and invokes only the
retrieval stage. It covers 56 inputs with three replays each. Canonical JSON
comparison recursively removes only \texttt{elapsed\_seconds}; ordered
mechanisms and scores, injected entry IDs and content, repository-fetch
records, and internal hand-off fields remain. All 56 replay groups are exact,
and all reproduce the corresponding recorded public Stage~2 payload. The
replay script emits a per-input CSV and a JSON report under schema
\texttt{malpr-stage2-fixed-input-replay-v1}. Across the 56 inputs, Stage~2 selected and injected 165 mechanisms in
total, two or three per run. This conditional result is not included in the
end-to-end cross-run guarantee.

\paragraph{Stage~2 stability under a varying upstream query.}
Fixed-input determinism is separate from stability when the model-produced Stage~1 query
changes. In the 14 paired Pool~A reruns per backend, the Stage~1 query hash changed in every
pair (28/28 overall), while the recorded Stage~0 evidence pack and its file set matched in
14/14 pairs for each backend. Table~\ref{tab:stage2-stability} measures how much of the
Stage~1 variation survives into the selected top-$\leq3$ knowledge-base mechanisms.

\begin{table}[htbp]
\centering
\caption{Stage~2 KB-selection stability over paired Pool~A reruns. ``Set'' is
exact selected-set equality; $\bar J$ is mean set Jaccard.}
\label{tab:stage2-stability}
\small
\begin{tabular}{@{}lrrr@{}}
\toprule
Backend & Same set & Same top-1 & $\bar J$ \\
\midrule
GPT-5.5 & 12/14 (86\%) & 13/14 (93\%) & .94 \\
DeepSeek & 7/14 (50\%) & 12/14 (86\%) & .75 \\
\bottomrule
\end{tabular}
\end{table}

\section{Uncovered Language Slice}
\label{app:lua}
Lua is absent from Stage~0's configured extension set, but we retain all four
APISIX cases rather than exclude this uncovered slice. Outcomes are mixed; the
real-IP case forms the right hypothesis with repository context but never
validates Admin-API reachability.

\begin{table}[htbp]
\centering
\caption{Diagnosis score, $(I+E)/2$, on four Lua APISIX cases.}
\label{tab:lua}
\small
\begin{tabular}{@{}lrrrrrrrr@{}}
\toprule
& \multicolumn{4}{c}{GPT-5.5} & \multicolumn{4}{c}{DeepSeek} \\
\cmidrule(lr){2-5}\cmidrule(lr){6-9}
Case & E0 & E1 & E2 & E3 & E0 & E1 & E2 & E3 \\
\midrule
24112 (real-IP) & 0 & .5 & .5 & .5 & 0 & .5 & .5 & .5 \\
29266 (JWT secret) & 1 & 1 & 1 & 1 & 1 & 1 & 1 & 1 \\
46647 (issuer) & 0 & 0 & 0 & 0 & 0 & 0 & 0 & 0 \\
62232 (auth log) & 1 & 1 & 1 & 1 & 0 & 1 & 0 & 0 \\
\bottomrule
\end{tabular}
\end{table}

\end{document}